\documentclass[aps,prl,twocolumn,superscriptaddress,showpacs]{revtex4}
\usepackage{graphicx}
\usepackage{rotating,dcolumn}
\begin{document}

\title{Single domain magnetic helicity and triangular chirality
in structurally enantiopure Ba$_3$NbFe$_3$Si$_2$O$_{14}$}

\author{K. Marty}
\affiliation{Institut N\'eel, CNRS \& Universit\'e Joseph Fourier,
BP166, 38042 Grenoble, France}
\author{V. Simonet} \affiliation{Institut N\'eel, CNRS \& Universit\'e Joseph Fourier, BP166, 38042 Grenoble, France}
\author{E. Ressouche}
\affiliation{Institut Nanosciences et Cryogenie, SPSMS/MDN,
CEA-Grenoble, 38054 Grenoble, France}
\author{R. Ballou}
\affiliation{Institut N\'eel, CNRS \& Universit\'e Joseph Fourier,
BP166, 38042 Grenoble, France}
\author{P. Lejay}
\affiliation{Institut N\'eel, CNRS \& Universit\'e Joseph Fourier,
BP166, 38042 Grenoble, France}
\author{P. Bordet}
\affiliation{Institut N\'eel, CNRS \& Universit\'e Joseph Fourier,
BP166, 38042 Grenoble, France}

\date{\today}

\begin{abstract}
A novel doubly chiral magnetic order is found out in the
structurally chiral langasite compound
Ba$_3$NbFe$_3$Si$_2$O$_{14}$. The magnetic moments are distributed
over planar frustrated triangular lattices of triangle units. On
each of these they form the same triangular configuration. This
ferrochiral arrangement is helically modulated from plane to
plane. Unpolarized neutron scattering on a single crystal
associated with spherical neutron polarimetry proved that a single
triangular chirality together with a single helicity is stabilized
in an enantiopure crystal. A mean-field analysis allows to discern
the relevance on this selection of a twist in the plane to plane
supersuperexchange paths.
\end{abstract}


\maketitle

Chirality is the geometric property of an object according to
which this exists in two distinct enantiomorphic states that are
images of each other by space inversion but cannot be brought into
coincidence by direct Euclidian isometry, namely spatial proper
rotation and translation, eventually combined with time reversal
\cite{flack}. An example in magnetism is the left or right
handedness associated with the helical order of magnetic moments.
Such an order may emerge from spontaneous symmetry breaking in
systems with competing exchange interactions
\cite{villain,yoshimori} or from the instability of simple
magnetic orders with respect to Dzyaloshinskii-Moriya
anti-symmetric exchange interactions
\cite{dzyaloshinski,bak,nakanishi,cscucl3}. The parity symmetry is
globally restored in the case of a centrosymmetric structure by
the presence in the same crystal of equally populated domains of
opposite chirality states, which can be unbalanced only through
axial-polar, magnetoelastic or magnetoelectric, field couplings
\cite{siratori,plakhty}. A single domain should be selected in
non-centrosymmetric compounds where the parity symmetry is
explicitly broken \cite{gukasov,maleyev}. So far this was reported
only in the intermetallic compound MnSi \cite{shirane}. Another
example of magnetic chirality is the clockwise or anticlockwise
asymmetry associated with the triangular configuration of
frustrated magnetic moments on a triangular plaquette with
antiferromagnetic interactions. Actual magnets exhibiting this
triangular chirality \cite{grohol}, eventually coexisting with the
helical chirality \cite{kenzelmann}, are scarce and none was found
in a single domain chiral state. With edge-sharing plaquettes the
choice of a chirality state on one triangle fixes to opposite the
chirality state on the adjacent triangles \cite{kawachi}. No such
constraint exists in the kagom\'e lattice of corner-sharing
triangles. Anti-symmetric exchange interactions, allowed in this
case, can select a uniform distribution of a chirality state
\cite{elhajal}, but this is altered by low energy defects
\cite{chalker}, bringing about alternative chirality textures. We
hereafter report on an attractive material,
Ba$_3$NbFe$_3$Si$_2$O$_{14}$, where the two magnetic chiralities
coexist, are single domain, and are fixed with respect to the
structural chirality.

\begin{figure}[t]
\includegraphics[bb=30 550 595 790,scale=0.5]{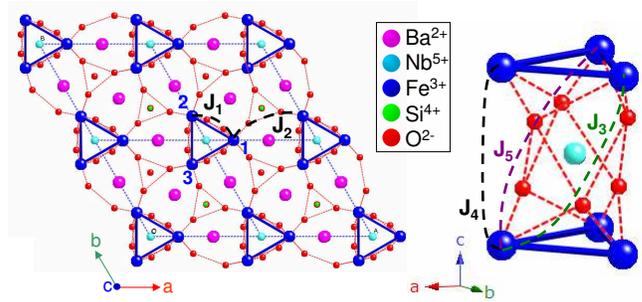}
\caption{(Color online) Ba$_3$NbFe$_3$Si$_2$O$_{14}$ crystal
structure as projected in the ({\it a}, {\it b}) plane and viewed
as stacked along the trigonal {\it c}-axis. The Fe$^{3+}$
triangles are shown with solid lines. The super and
supersuperexchange paths are depicted with short-dashed lines. The
$J_1$ to $J_5$ exchange links are schematized by long-dashed
lines.} \label{f.1}
\end{figure}

Ba$_3$NbFe$_3$Si$_2$O$_{14}$ crystallizes in the P321
non-centrosymmetric trigonal space group and is isostructural to
La$_3$Ga$_5$SiO$_{14}$, thus belonging to the so-called langasite
family. These formerly were studied for their piezoelectricity
\cite{iwataki} and their non-linear optical and electro-optical
properties \cite{xin}. A strong interest in their magnetic
behaviour was aroused more recently when it was realized that some
of them would materialize kagom\'e lattice of rare-earth cations
\cite{bordet} or, as in the present case, triangular lattice of
triangle units of transition metal cations \cite{marty}. As
schematized in Fig. \ref{f.1}, this lattice is coplanar to the
({\it a}, {\it b}) plane and formed in
Ba$_3$NbFe$_3$Si$_2$O$_{14}$ by the Fe$^{3+}$ ions with spin
S=5/2, which  are the only magnetic cations. Consecutive planes
are separated by layers containing Ba and Nb cations. The
Fe$^{3+}$ ions are tetrahedrally coordinated by oxygens anions,
which mediates the superexchange interaction within the triangles.
The magnetic interaction between spins belonging to different
triangles is mediated by two oxygens (supersuperexchange) in the
({\it a}, {\it b}) planes and also along the {\it c}-axis. It thus
is expected to be weaker than the intra-triangle interaction.

\begin{table}
\caption{Structural parameters of Ba$_3$NbFe$_3$Si$_2$O$_{14}$ as
refined from X-ray single crystal diffraction data at room
temperature. U$_{eq}$ is the isotropic displacement parameter in
\AA$^2$. The lattice parameters are $a$=$b$=8.539(1) \AA,
$c$=5.2414(1) \AA. The agreement factors are : R$_{wall}$ 1.69\%
and goodness of fit 1.19.\%.}
 \label{tab}
 \begin{tabular}{cccccccc}
 \hline\hline
 Atom   &  Wyckoff &  $x$    &  $y$      &  $z$      & U$_{eq}$\\
 \colrule
 Ba  &  3e &  0.56598(2) & 0 & 0 &  0.00859(4)\\
 Nb  &  1a  &  0 & 0 & 0 & 0.00766(6)\\
 Fe  &  3f   &  0.24964(4) & 0 & 1/2 & 0.00776(7)  \\
 Si  &  2d  &  2/3 & 1/3 & 0.5220(1) & 0.0063(1) \\
 O(1)  &  2d  &  2/3 & 1/3 & 0.2162(4) & 0.0106(4) \\
 O(2)  &  6g  &  0.5259(2) & 0.7024(2) & 0.3536(3) & 0.0118(4)\\
 O(3)  &  6g  & 0.7840(2) & 0.9002(2) & 0.7760(3) & 0.0164(4)\\
 \hline\hline
 \end{tabular}
\end{table}

Powders of Ba$_3$NbFe$_3$Si$_2$O$_{14}$ were synthesized by solid
state reaction from stoichiometric amounts of Nb$_2$O$_3$,
Fe$_2$O$_3$, SiO$_2$ oxides and BaCO$_3$ barium carbonate, at 1150
$^{\circ}$C in air, within an alumina crucible. The reagents were
carefully mixed and pressed to pellets before annealing. The phase
purity was checked by X-ray powder diffraction. Single-crystals
were grown by the floating-zone method in an image furnace
\cite{bordet}. Small fragments extracted from these were used to
investigate the crystal structure on a BrukerÐNonius kappaCCD
x-ray diffractometer using the Ag K$\alpha$ radiation. The
anomalous part of the scattering function allowed us to infer the
crystal chirality. The one associated with the atomic positions
reported in table \ref{tab} is called left-handed, in view of the
anti-trigonometric twist of the exchange paths around the {\it
c}-axis (see Fig. \ref{f.1} and focus at the $J_5$ exchange path,
which then is the dominant interplane interaction). Note that
there are three Fe$^{3+}$ ions per unit cell (labeled in
Fig.\ref{f.1}) deduced from each other by the 3-fold symmetry.

\begin{figure}[t]
\includegraphics[bb=20 550 572 814, scale=0.45]{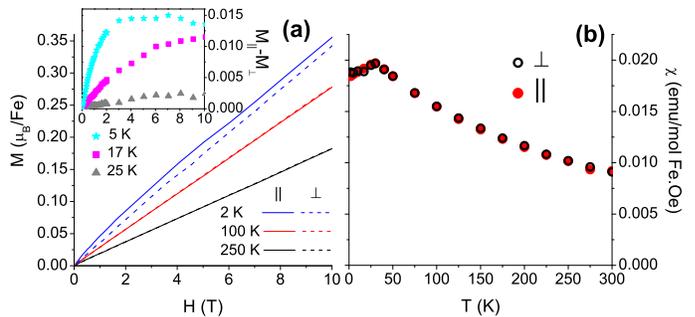}
\caption{(Color online) a) Magnetization versus magnetic field
with the field applied $\parallel$ or $\perp$ to the {\it c} axis.
Inset: difference between $\parallel$ and $\perp$ curves below
$T_N$. b) High field susceptibility versus temperature for both
$\parallel$ and $\perp$ orientations.} \label{f.2}
\end{figure}

Magnetization measurements were performed on a
Ba$_3$NbFe$_3$Si$_2$O$_{14}$ single-crystal from 2 to 300 K under
magnetic fields up to 10 T on a purpose-built magnetometer. The
isotherms of the magnetization $M$, shown in Fig.\ref{f.2}a, are
linear and independent on the applied field orientation, parallel
($\parallel$) or perpendicular ($\perp$) to the {\it c} axis, at
least down to the ordering transition temperature. No significant
anisotropy is thus detected in the paramagnetic phase as expected
for an Fe$^{3+}$ ion with a spin $S$=5/2 and no orbital
contribution. At low temperatures, the high field linear part of
the magnetization isotherms remains identical for both
orientations. The associated susceptibility is shown in Fig.
\ref{f.2}b where a cusp at T$_N\approx$ 27 K signals the
transition towards a magnetic order, also pointed out from a sharp
peak in the specific heat \cite{marty}. Below T$_N$, the
$\parallel $ magnetization isotherms, as opposed to the $\perp$
ones,  exhibit a slight curvature revealing the rise of a small
magnetic component along the trigonal {\it c} axis. Its value is
estimated $\approx$ 0.014 $\mu_B$/Fe at 2 K from the difference
between the $\perp$ and $\parallel$ magnetizations (see inset of
Fig. \ref{f.2}a).  Above 100 K, the magnetic susceptibility is
fitted by a Curie-Weiss law $\chi=C/(T-\theta)$, with a Curie
constant $C$ corresponding to the effective moment $\mu_{\rm
eff}=g\sqrt{S(S+1)}$=5.92 $\mu_B$ of an Fe$^{3+}$ ion and a
Curie-Weiss temperature $\theta$ = -174$\pm $4 K. The latter
indicates predominant antiferromagnetic exchange interactions and
suggests magnetic frustration or dimensional reduction since the
magnetic ordering occurs at a much lower temperature. A
coexistence of phase in a narrow range of temperature around
T$_N$ was observed by $^{57}$Fe M\"ossbauer spectroscopy,
implying that the magnetic transition would be first order \cite{marty}.

Neutron diffraction experiments were carried out at the Institut
Laue Langevin (ILL), first on powder samples using the D1B
diffractometer \cite{marty} then on a single crystal using the D15
4-circles diffractometer. Magnetic Bragg peaks emerge below $T_N$.
They can be indexed using the propagation vector (0, 0, $\tau$)
with $\tau$ close to 1/7 \cite{marty}. The magnetic structure, as
determined from refinement of the powder and single-crystal
diffractogramms, consists in magnetic moments lying in the ({\it
a}, {\it b}) planes at 120$^{\circ}$ from each other within each
triangle. This magnetic arrangement is accordingly repeated from
cell to cell in the ({\it a}, {\it b}) planes. On moving along the
{\it c} axis the spins rotate to form an helix of period $\approx$
7 lattice parameters. At 2 K, the fitted value of the magnetic
moment is $\approx$ 4 $\mu_B$, instead of the expected 5 $\mu_B$
for an Fe$^{3+}$ ion. This reduction may result from spin transfer
to the oxygen ions \cite{Fe3O4}. The obtained magnetic structure
is consistent with symmetry analysis, which yields three
1-dimensional irreducible representations, leading to the
prediction of the two 120$^{\circ}$ structure with opposite
triangular chirality and a structure with ferromagnetic planes
helically propagating along the  {\it c} axis. Whereas powder
diffraction is insensitive to the triangular chirality and
helicity, single-crystal diffraction can bring additional
information on this issue. For clarity in the following and using
the trigonometric convention, the term triangular chirality will
refer to the sense of rotation of the 3 spins within the triangle
while going trigonometrically from one corner of the triangle to
the others. The term helicity will refer to the sense of rotation
of the spins in an helix along the direction of the propagation
vector.

\begin{figure}[t]
\includegraphics[bb=0 150 814 572,scale=0.3]{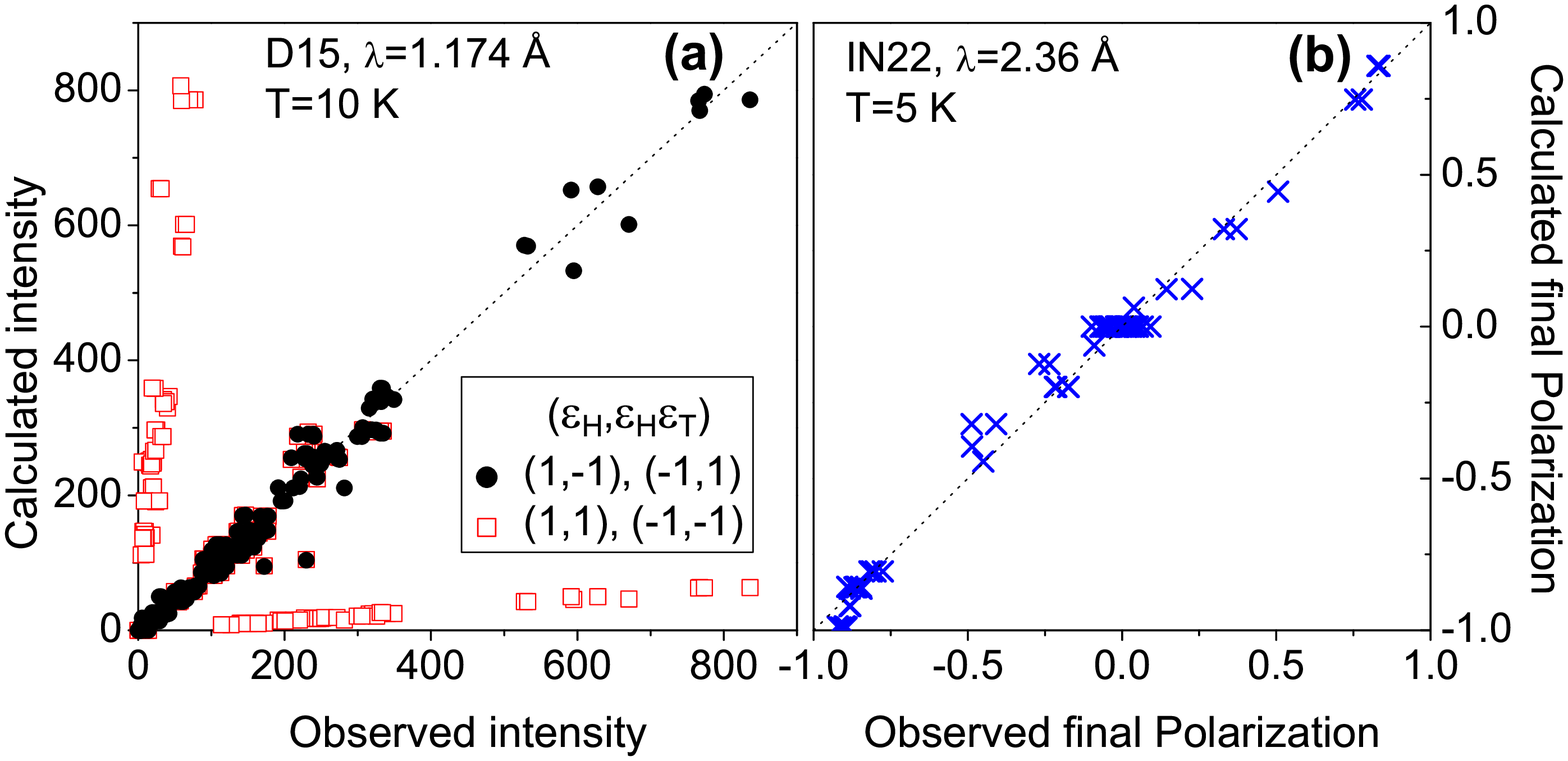}
\caption{(Color online) (a) Single crystal diffraction: calculated
versus observed integrated magnetic intensities for different
($\epsilon_H$, $\epsilon_H\epsilon_T$) pairs (see text). (b)
Spherical polarization analysis: calculated versus observed
final polarization for 8 magnetic Bragg peaks. The calculated
values are those obtained from a refinement of the distribution
of domains, which yields the best agreement with the observed
values for a single domain of helicity.} \label{f.3}
\end{figure}

The expression of the magnetic moments in the $n^{\rm{th}}$ unit
cell for a perfect helix propagating along the {\it c} axis with 3
Bravais lattices (labeled $i$=1,2,3 in Fig.\ref{f.1}) is written
as $\vec{m}_i(\vec{R}_n)=m\cos(\vec{\tau}\cdot\vec{R}_n+\epsilon_T
\Phi_i)\vec{u}+\epsilon_H
m\sin(\vec{\tau}\cdot\vec{R}_n+\epsilon_T \Phi_i)\vec{v}$, where
$\vec{u}$ and $\vec{v}$ are orthonormal vectors in the ({\it a},
{\it b}) plane, forming a right handed set with the {\it c} axis,
and $\Phi_i$ is the phase of the $i^{\rm{th}}$ Bravais lattice.
These are accessible from the diffraction data only through the
modulus $\vert \Phi_{i+1}-\Phi_i \vert$ of their differences.
$\epsilon_H=\pm1$ determines the helicity
($\vec{m}_i(\vec{R}_n)\wedge\vec{m}_i(\vec{R}_n+\vec c)=
m^{2}sin(\tau) ~ \epsilon_H ~ {\vec{c} / \vert \vec{c} \vert}$)
and $\epsilon_H\epsilon_T=\pm1$ the triangular chirality ($\sum_i
\vec{m}_i(\vec{R}_n)\wedge\vec{m}_{i+1}(\vec{R}_n)= (3\sqrt
2/2)m^{2} ~ \epsilon_H\epsilon_T ~ {\vec{c} / \vert \vec{c}
\vert}$). Only two pairs among the four possible magnetic
chirality states, ($\epsilon_H$, $\epsilon_H\epsilon_T$)=(1,-1)
and (-1,1), are found compatible with the unpolarised neutron's
single-crystal data refined for a left-handed structural chirality
(black circles in Fig. \ref{f.3}). An additional experiment on
single-crystal using polarized neutrons with spherical
polarization analysis was performed using the CRYOPAD device on
the IN22 spectrometer at the ILL. This allows one to measure the
three orthogonal components of the polarization vector of the
neutron beam after scattering by the sample whatever the
polarization of the incoming neutron beam. It however suffices to
choose three orthogonal orientations of this initial polarisation
to get all the accessible informations at a scattering vector
($Q_h$, $Q_k$, $Q_l$). This leads to nine independant data, giving
the components $\overline{P}_{i,j}$ ($i$,$j$=X,Y,Z with X
$\parallel$ scattering vector and Z $\perp$ scattering plane) of
the so-called polarization matrix $\overline{P}$ \cite{pol}. This
was measured on 8 magnetic peaks of the type (-1, 2, $l\pm \tau$)
and (1, -2, $l\pm \tau$) with $l\in$[0,3]. It is found out from
the formalism \cite{blume} that the polarization matrices
$\overline{P}$ are only sensitive to the helicity $\epsilon_H$,
with the components $\overline{P}_{Y,X}$ and $\overline{P}_{Z,X}$
being proportional to the associated distribution of helicity
domains. A fit of our data with respect to domain proportions
systematically leads to a single helicity in the crystal (see Fig.
\ref{f.3}b). This, in turn, indicates the selection of one
($\epsilon_H$, $\epsilon_H\epsilon_T$) pair, to agree with the
unpolarized diffraction data, thus the selection of the unique
associated triangular chirality.

\begin{figure}[t]
\includegraphics[bb=20 270 595 800,scale=0.45]{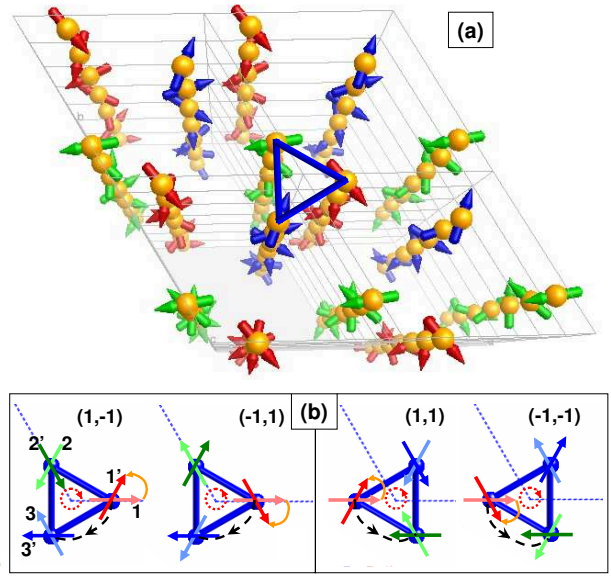}
\caption{(Color online) (a) Perspective view of the magnetic
structure with different colors for the three Bravais lattices.
(b) Representation of the magnetic structures associated to the 4
possible ($\epsilon_H$,$\epsilon_H\epsilon_T$) pairs. The light
colored moments lie in one layer and the darker colored ones in
the next layer along the {\it c} axis, an orange curved arrow
defines the helicity. The red arrowed circle materializes the
triangular chirality. The structural chirality is represented by a
diagonal exchange between the two layers (dashed arrow path).}
\label{f.4}
\end{figure}

A zero temperature mean-field analysis was undertaken to get more
insights about the magnetic structure and to relate it to the
crystal structure \cite{bertaut}. A set of five exchange
interaction parameters were considered in the Heisenberg
Hamiltonian $\mathcal{H}=-\frac{1}{2}\sum_{i,j}
J_{k}\vec{S_i}\cdot\vec{S_j}$, namely $J_1$ the intra-triangle first
neighbor interactions, $J_2$ the inter-triangle second neighbor
interactions in the ({\it a}, {\it b}) plane and $J_3$ to $J_5$
the inter-triangle interactions of adjacent planes (see Fig.
\ref{f.1}). These last three supersuperexchange paths are
non-equivalent. As from the crystal geometry, it appears that the
strongest one (shorter bond lengths and bond angles closer to
180$^{\circ}$ \cite{goodenough}) would be the diagonal $J_5$
interaction. It links spins screw-like along the {\it c} axis,
trigonometrically (anti-trigonometrically) for right-handed
(left-handed) structural chirality. With antiferromagnetic $J_1$
and $J_2$ and null inter-plane interactions ($J_3$ to $J_5$=0),
the diagonalization of the Fourier transform of the interaction
matrix yields three solutions with zero propagation vector : a
less favored ferromagnetic order and two degenerate 120$^{\circ}$
spin structures with opposite triangular chirality. With
additional weak inter-plane interactions where one diagonal
interaction prevails over the other two, a helical modulation is
generated along the {\it c} axis and the degeneracy of the two
120$^{\circ}$ spin configurations is lifted.
The favored solution for left-handed structural chirality then
corresponds to the two ($\epsilon_H$, $\epsilon_H\epsilon_T$)
pairs found with unpolarized neutrons single-crystal diffraction.
Inverting the structural chirality we get the two other
($\epsilon_H$, $\epsilon_H\epsilon_T$) pairs (see Fig.\ref{f.4}b).

Additional understanding can be gained by simple geometrical
considerations. Given the 120$^{\circ}$ spin structure in the
triangles, if one considers only one predominant diagonal
antiferromagnetic interaction between adjacent layers, one atom of
a given triangle will be anti-aligned with the atom of the upper
triangle in the diagonal direction ({\it e.g.} atoms 1 and 3' in
Fig. \ref{f.4}b). This will result in a rotation by 60$^{\circ}$
of the spins along the {\it c} axis ({\it e.g.} from atom 1 to
atom 1') leading to a propagation vector (0, 0, 1/6), close to the
value determined in Ba$_3$NbFe$_3$Si$_2$O$_{14}$. The departure
from 1/6 can be ascribed to the contribution of the other two
inter-plane interactions. Therefore, the observed magnetic
arrangement is well described by the 5 considered exchange
interactions. This is unlike MnSi or CsCuCl$_3$, where the helices
are generated from Dzyaloshinskii-Moriya interactions
\cite{bak,cscucl3}. The mean field calculation though can only
access the modulus of the phase difference and is energetically
favorable to two ($\epsilon_H$,$\epsilon_H\epsilon_T$) solutions.
As polarized neutrons have shown that one of this solution only is
actually observed, the origin of this ultimate selection is still
unclear. It could be due to the Dzyaloshinskii-Moriya
antisymmetric interaction, allowed in this compound. Another hint
for the presence of this interaction is the small ferromagnetic
component observed along the {\it c} axis in the magnetization
isotherms (see Fig. \ref{f.2}a).

It is clear from our results that Ba$_3$NbFe$_3$Si$_2$O$_{14}$
could raise a strong interest in the field of multiferoism.
Indeed, although unexpected when the propagation vector is
perpendicular to the helix plane \cite{katsura,mostovoy}, the
onset of an electrical polarisation was observed at the magnetic
ordering temperature in another trigonal helically stacked
triangular antiferromagnets with a 120$^{\circ}$ spin structure
\cite{kenzelmann}. Two chiral magnetic domains coexist in this
centrosymmetric compound. From phenomenological symmetry
arguments, the amplitude of the electrical polarization was
related to the unbalance between these two domains. If this
argument holds, the effect should be maximum in
Ba$_3$NbFe$_3$Si$_2$O$_{14}$ since the chirality is single domain.
To probe this, measurements of the dielectric constant were
performed. An anomaly was observed at the onset of the magnetic
order. This will be completed with electric polarisation
measurements and reported elsewhere.

In conclusion, the Fe-langasite Ba$_3$NbFe$_3$Si$_2$O$_{14}$
provides the first evidence of a totally chiral state from the structural
point of view where it manifests itself by the twist of the exchange
paths and from the magnetic point of view where two different kinds
of chiralities, within the triangles and along the helices, coexist and
interplay.

\acknowledgments This work was financially supported by the ANR
06-BLAN-01871. We would like to thank B. Canals and L.-P. Regnault
for fruitful discussions and for the help during the polarized
neutron experiment for the latter.

\end{document}